\documentstyle[aps,twocolumn,pra,psfig]{revtex}

\begin{document}

{\bf Comment on ``Demonstration of the Casimir Force in the 0.6 to 6 $\mu$m 
Range''} 

In a recent Letter \cite{Lamoreaux97}, Lamoreaux reports a measurement 
of the Casimir force for distances in the 0.6 to 6 $\mu$m range. The force has 
been measured between a flat and a spherical plate. Both plates are coated with 
a layer of Cu, covered with an additional 0.5 $\mu$m thick layer 
of Au. The author compares his experimental data to theoretical predictions 
\cite{Lamoreaux98} and reports an agreement at 5\% between theory and experiment
if a pure Cu surface is assumed. Since his theoretical evaluation 
\cite{Lamoreaux99} gives very different 
values for the Casimir force between Au surfaces and Cu surfaces, there
results a net discrepancy between expected and experimentally observed values.

We have recently recalculated the Casimir force between metallic mirrors and
obtained results differing significantly from \cite{Lamoreaux98,Lamoreaux99}, in 
particular 
for Au mirrors. 
Details about the evaluation procedure, the interpolation and 
extrapolation of optical data, and the numerical integration techniques
are given in \cite{Lambrecht00}. Here, we restrict our attention on the 
Au/Cu problem underlined by Lamoreaux.

\begin{figure}[tbh]
\centerline{\psfig{figure=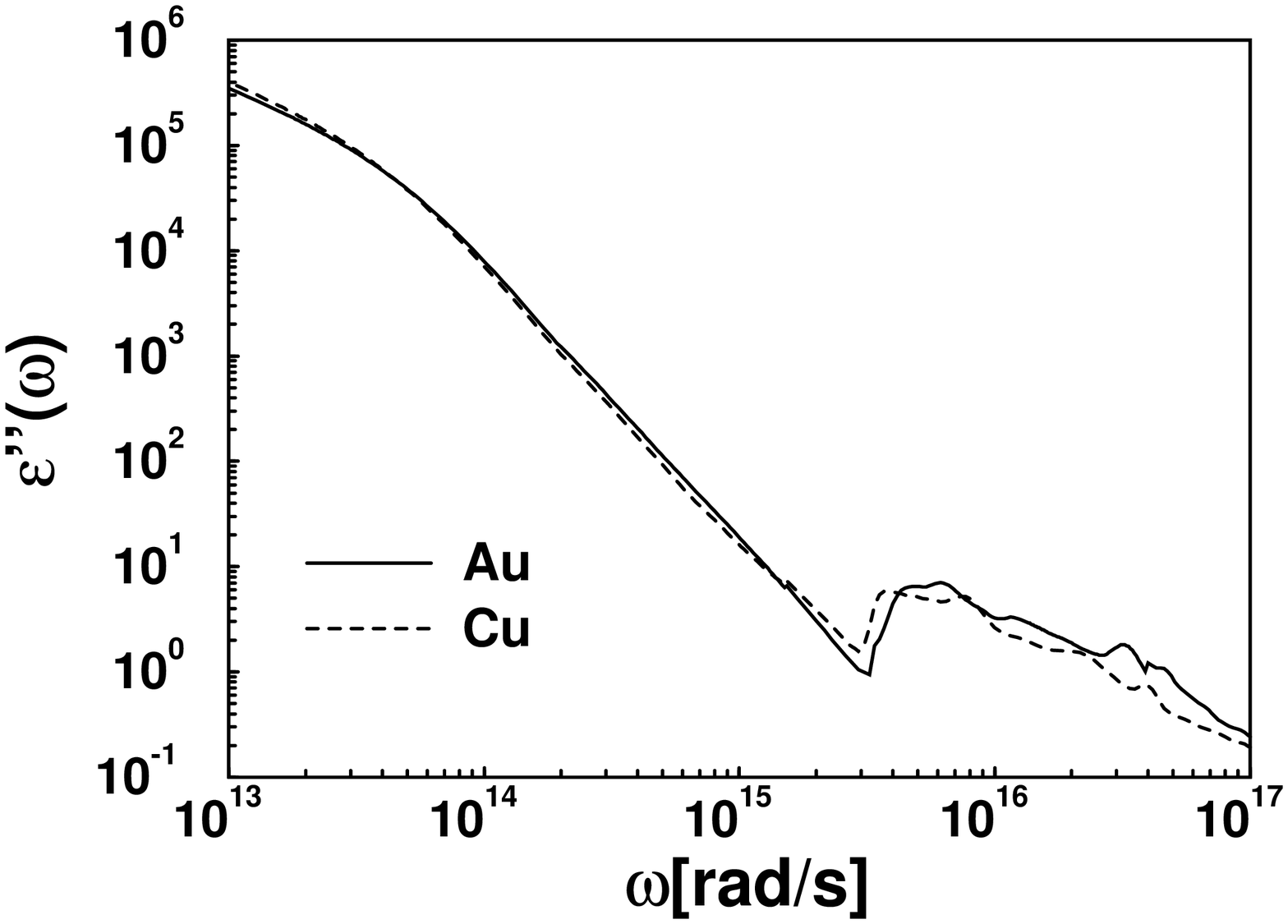,width=7cm}}
\centerline{\psfig{figure=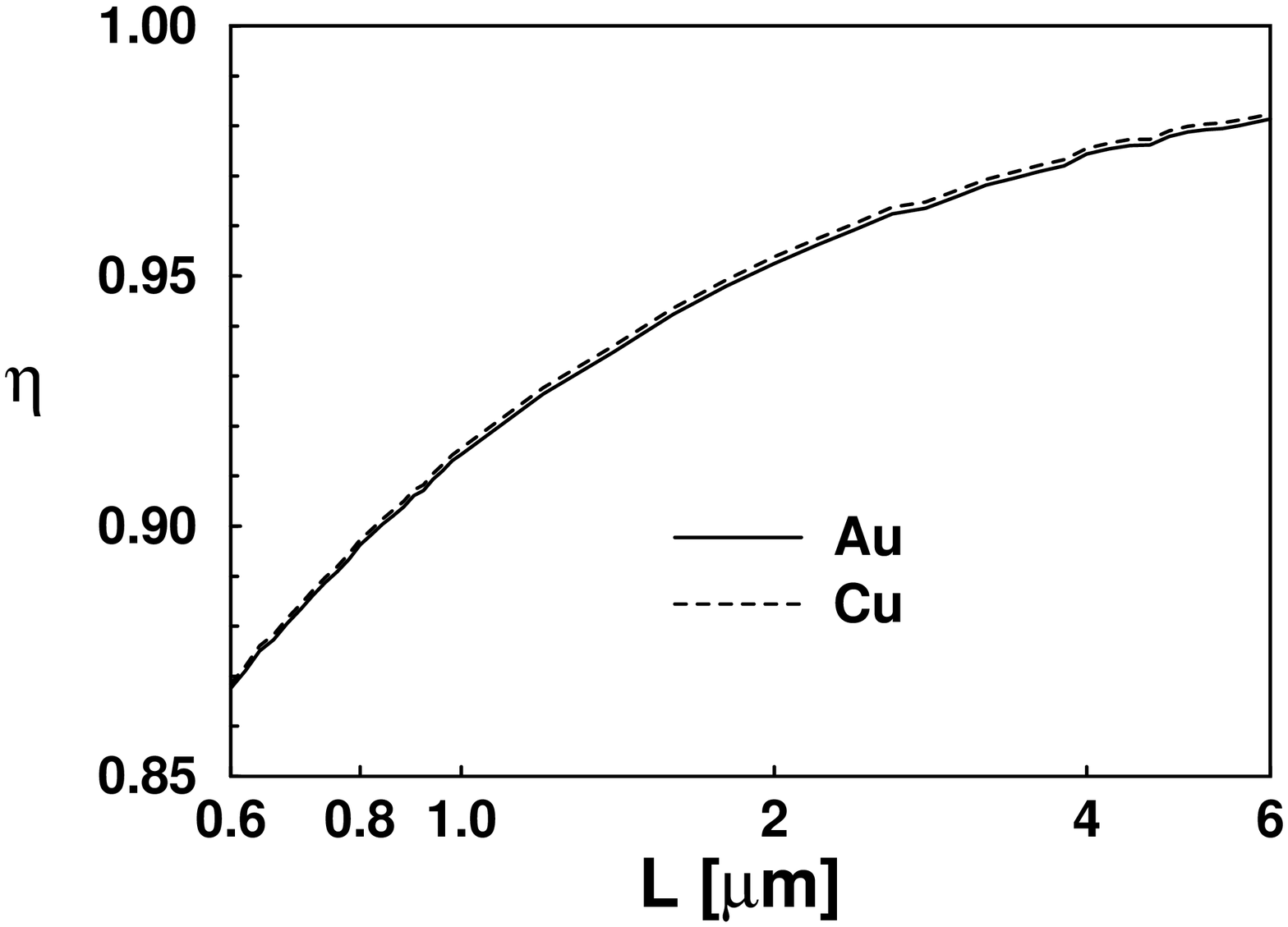,width=7cm}}
\caption{The imaginary part of the dielectric constant as function of 
frequency (upper graph) and the reduction of the Casimir energy between 
plane metallic reflectors with respect to plane perfect
mirrors as a function of distance (lower graph) for Au (solid  line) and Cu 
(dashed line).}
\label{fig}
\end{figure}

The upper graph of figure \ref{fig} shows the imaginary part of the dielectric 
constant $\varepsilon ^{\prime \prime} (\omega )$ as a function of frequency 
$\omega$ for Au and Cu. All optical data are taken from 
\cite{Palik}. At low frequencies they are extrapolated by a Drude model which is 
consistent with present theoretical knowledge of optical properties of metals 
and, at the same time, fits quite nicely higher frequency optical data. 
Since the optical response functions are very similar for Au and Cu,
the Casimir forces evaluated from these functions are expected to be 
nearly equal.

In the experiment, the Casimir force is measured in the plane-sphere geometry. 
Theoretically it is evaluated by using the proximity force 
theorem. We do not discuss here the validity of this approximation but
focus our attention on the effect of finite conductivity. 
We calculate the reduction factor $\eta$ (notation of \cite{Lamoreaux98}; 
notation 
$\eta_E$ in \cite{Lambrecht00}) of the force in the plane-sphere 
geometry as the reduction factor of the energy evaluated in the 
plane-plane configuration.
The frequency dependent reflection coefficients are derived from the dielectric 
constant, using causality relations, and $\eta$ is then deduced
through numerical integrations. 

The lower graph of figure \ref{fig} shows 
$\eta$ for Au and Cu with, as expected, equal values at better than 1\% in the 
range of 
distances studied in the experiment. This contradicts theoretical values 
obtained by Lamoreaux. For Au at $0.6~\mu$m our value 
$\eta = 0.87$ exceeds by 12\% the value $\eta = 0.78$ given in 
\cite{Lamoreaux98}, while 
at the same distance the values for Cu are compatible within 2\%.

This result clears up the Au/Cu discrepancy pointed out in 
\cite{Lamoreaux98}. 
Besides this specific difficulty, more work is needed, on both the experimental 
and theoretical side, to reach an accurate agreement between theoretical
expectation and experimental measurements of Casimir force (see 
\cite{Lambrecht00}
and references therein).

\vspace*{0.5cm}
\noindent
Astrid Lambrecht and Serge Reynaud\\
\hspace*{4.8mm}Laboratoire Kastler Brossel \cite{LKB}\\
\hspace*{4.8mm}Campus Jussieu, case 74 \\
\hspace*{4.8mm}75252 Paris Cedex 05, France

\vspace*{0.5cm}
\noindent
PACS numbers: 12.20 Fv, 07.07 Mp, 03.70 +k

\end{document}